 \documentclass[final,3p,times]{elsarticle}

 \usepackage{graphicx}

\usepackage{amssymb}
 \usepackage{amsthm}
 \usepackage{amsmath}
\usepackage{color}

\journal{Physics Letters A}

\begin{document}

\begin{frontmatter}



\title{Fermionic van Hemmen Spin Glass Model with a Transverse Field }


\author[uff]{S. G. Magalhaes}\ead{ggarcia@ccne.ufsm.br}\fntext[ufsm]{Tel.: +55 (055) 3220
9521}
\author[ufsm]{F. M. Zimmer}
\author[ufsm]{C. V. Morais}

\address[uff]{Instituto de Fisica, Universidade Federal Fluminense, 24210-346 Niter\'oi,  RJ, Brazil }
\address[ufsm]{Departamento de Fisica, Universidade Federal de Santa Maria,
97105-900 Santa Maria, RS, Brazil}

\begin{abstract}

In the present work  it is studied the fermionic van Hemmen model for the spin glass (SG) with
a transverse magnetic field $\Gamma$. In this model, the spin operators are written as a bilinear
combination of fermionic operators, which allows  the analysis of the interplay between
charge and spin fluctuations in the presence of a quantum spin flipping mechanism given by
$\Gamma$. The problem is expressed in the fermionic path integral formalism. As results, magnetic
phase
diagrams  of temperature versus the ferromagnetic interaction are obtained for several values
of chemical potential $\mu$ and $\Gamma$. The $\Gamma$ field suppresses the magnetic orders. The
increase of $\mu$ alters the average occupation per site that affects the magnetic phases. For
instance, the SG and the mixed SG+ferromagnetic phases are also suppressed by $\mu$. In addition,
$\mu$ can change the nature of the phase boundaries introducing a first order transition.
\end{abstract}


\end{frontmatter}



It is well established that disorder can be a source of non-trivial effects in condensed 
matter physics. From Anderson localization to spin glass, there are plenty of examples supporting
such role. As one more recent example there are indications that the presence of disorder is
responsible by several complex effects in strongly correlated systems as, for instance, manganites
or Cerium compounds (see Ref. \cite{dagotto} and references therein). The important point is that
disorder can affect the charge sector as well as the spin one in such way that it becomes a constant
cause of new interesting problems. From the theoretical point of view, the task of facing this kind
of problems is surely a source of new approaches.

Recent experimental findings in two specific Cerium compounds can illustrate the last comment in the
previous paragraph. $CePd_{1-x}Rh_x$ \cite{steglich} and $CeNi_{1-x}Cu_x$ \cite{marcano} are
physical systems in which disorder combined with RKKY and Kondo interaction is responsible by the
onset of a complex scenario. Their phase diagrams display a glassy state, ferromagnetism and a
region dominated by the Kondo effect. For $CeNi_{1-x}Cu_x$, several works have used the so called
Kondo-Ising lattice (KIL) model \cite{coqblin} to reproduce the experimental phase diagram of that
compound. The Ising part of the model represents an intersite interaction between localized f-spin
operators. It is precisely in the spin-spin Ising coupling $J_{ij}$ that various types of randomness
have been tested trying to represent the effects of the disorder in the $CeNi_{1-x}Cu_{x}$
\cite{alba1, magalhaesepjb, alba2, magalhaesprb1, magalhaesphysicab}. For instance, the random
coupling $J_{ij}$ has been chosen as the Sherrington-Kirkpatrick (SK) model \cite{sk} or as the
Hopfield model \cite{hebb}. However, none of the choices previously mentioned were entirely
satisfactory to describe how 
the glassy state in the
$CeNi_{1-x}Cu_{x}$ is replaced by an inhomogeneous ferromagnetism when the temperature is decreased
\cite{marcano}.

Very recently, a new approach has been proposed for the problem discussed above, this approach uses
the random coupling $J_{ij}$ as that one introduced in the van Hemmen (vH) model  \cite{vh}. This
model was originally conceived to study the spin glass (SG) problem with classical Ising spin
variables. The results indicated that the vH type of randomness in the KIL model is more adequate
than those ones used in SK or Hopfield models to reproduce some aspects of the experimental phase
diagram of the $CeNi_{1-x}Cu_{x}$ \cite{magalhaesprb2}. For instance, the previously mentioned
ferromagnetic phase appears below the spin glass one.

The improvement obtained using the coupling $J_{ij}$ given as the vH model to study the
$CeNi_{1-x}Cu_{x}$ suggests the usefulness of this kind of randomness could be more general. In
other words, it could also be used when it is present in other quantum process than the Kondo
effect. In this case, models with one type of fermions could be built by adding to the vH Ising
term, for instance, a hopping, a pairing coupling \cite{magalhaesprb5} or a transverse magnetic
field. Nevertheless, surprisingly little consideration has been given to a fermionic version of the
vH model. In fact, even less consideration is given to the fermionic vH model when a term capable to
introduce quantum dynamics in the problem is added to this model. In fact, it lacks an investigation
of the interplay of charge and spin fluctuations within a fermionic version of the vH model even for
a simple quantum process provided by a transverse magnetic field. 

The purpose of the present work is to study the vH model in the presence of a transverse field
$\Gamma$ in a fermionic formulation which means that the Ising spin operators are written as
bilinear combinations of creation and destruction 
fermionic operators. For this particular
representation of the spin operators, the natural tool to obtain the thermodynamics is the
functional integral formalism with Grassmann fields. A peculiarity of this representation is that
$\hat{S}^{z}_{i}$ acts in a space with four eigenstates. Among them, there are two which are
nonmagnetic \cite{albaphysicaa}. Therefore, we calculate the grand partition function for two
situations. In the first one, here called $4S$ fermionic van Hemmen ($4S$-FvH) model, the four
states are preserved. For this particular model, the chemical potential $\mu$ appears as an
important parameter in the problem. In the second one, the $2S$ fermionic van Hemmen ($2S$-FvH)
model, introduces a restriction, what allows considering only the two magnetic states
\cite{albaphysicaa,wiethege}. One of the most important advantages of the vH model in both versions
4S and 2S is that the disorder can be treated without the use of the replica method. Therefore, the
subsequent difficulties associated with this technique \cite{binder} can be avoided in the present
work (see discussion below).  However, the problem is still treated within the static approximation
(SA) in which the time fluctuations are neglected \cite{bray}.

Indeed, the fermionic representation of Ising spin operators for the study of the SG problem is well
established. For the SK type of randomness, the so called fermionic Ising spin glass (FISG) model
\cite{wiethege,alba3,oppermann1} shows an interesting characteristic. Charge and spin correlation
functions are connected. The average occupation of fermions per site and the replica diagonal
element of the SG order parameter matrix are directly related. This relationship implies that, for
instance, the nature of the PM/SG phase boundary can be modified by variations of chemical potential
with the onset of a tricritical point \cite{oppermann2}. The presence of an additional transverse
magnetic field $\Gamma$ in the FISG model also introduces important changes in the PM/SG phase
boundary. The first order part of that phase boundary is diminished when $\Gamma$ increases
\cite{magalhaesprb3}. Consequently, the position of the tricritical point is also affected.
Therefore, one could expect that in the 4S-FvH model, variations of $\mu$ could also bring important
consequences on the location and nature of phase boundaries.

It should be remarked that the location of a first order PM/SG phase boundary is not a trivial task
for models in which the replica method is used as the FISG one \cite{magalhaesprb3,magalhaesprb4}.
This difficulty has already been a subject of controversy for the classical Ghatak-Sherrington model
(see, for instance, Refs. \cite{ghatak,salinas}). It is clear that this kind of controversy would
not be needed in the FvH model. On the other hand, vH model (classical or fermionic) 
has
a
shortcoming. It lacks to present a multiplicity of metastable-states \cite{choy}.  Nevertheless, the
vH model produces some results which are close to those ones found experimentally in SG physical
systems (see Ref. \cite{vh1}). Thus, the proper understanding of the vH model in a fermionic
formulation can be important not only as a SG problem itself but also as something useful for
applications as pointed out above and very recently discussed in Ref. \cite{magalhaesprb2}.

It should also be mentioned that the vH model with an additional transverse magnetic field $\Gamma$
has already been studied. However, in a formulation where the classical Ising spin variables were
replaced by Pauli matrices \cite{jricardo1}. In this formulation one can observe important
consequences for the phase diagram produced by the presence of $\Gamma$. The obtained phase diagram
shows that the increase of $\Gamma$ destroys the SG, the ferromagnetic (FM) and the mixed (FM+SG)
phases found in the classical phase diagram. Nevertheless, this kind of formulation is not able
to capture
the consequences for the phase diagram when both charge and spin fluctuations are affected by
$\Gamma$.

This paper has the following structure. In the second section, the model is introduced and the grand
canonical potential is derived. In the third section, several phase diagrams are constructed with
particular attention for the effects of $\Gamma$ and $\mu$ by solving the equations for the order
parameters. The last section is dedicated to the conclusions.

\section{Model}

The Hamiltonian of the fermionic van Hemmem (FvH)
model with a magnetic transverse field $\Gamma$ is
given by
\begin{equation}
 H = - \frac{2J_0}{N}\sum_{i\neq j}\hat{S}_{i}^{z}\hat{S}_{j}^{z}- 2\sum_{i\neq
j}J_{ij}\hat{S}_{i}^{z}\hat{S}_{j}^{z}-2\Gamma\sum_{i} \hat{S}_{i}^{x}
\label{ham}
\end{equation}
with the operators
\begin{equation}
 \hat{S}_{i}^{z}=\frac{1}{2}[c_{i\uparrow}^{\dagger}c_{i\uparrow}
- c_{i\downarrow}^{\dagger}c_{i\downarrow}],
\ \ \hat{S}_{i}^{x}=\frac{1}{2}[c_{i\uparrow}^{\dagger}c_{i\downarrow}+
c_{i\downarrow}^{\dagger}c_{i\uparrow}]
\end{equation}
and the random coupling $J_{ij}$ given by
\begin{equation}
 J_{ij}=\frac{J}{N}[\xi_{i}\eta_{j}+\xi_{j}\eta_{i}],
\label{Jij}
\end{equation}
where $\xi_{i}$ and $\eta_{i}$ are random variables which follow a bimodal distribution
\begin{equation}
 P(\xi_{i})=\frac{1}{2}[\delta(\xi_{i}-1) + \delta(\xi_{i}+1)].
\label {prob}
\end{equation}

In the present work the partition function is 
obtained within the Lagrangian path integral formalism.  
In that case, the spin operators are represented as bilinear combinations of 
Grassmann fields ($\varphi,~\varphi^{*}$).
Particularly, the $2S$-FvH model admits only magnetic states.  
This restriction for the corresponding partition function is obtained by using the Kronecker
$\delta$ function ($\delta(\hat{n}_{i_{}\uparrow}+ \hat{n}_{i_{}\downarrow}-1)$ $= 
\frac{1}{2\pi} \int_{0}^{2\pi} dx_{i_{}}$ $e^{ix_{i}(\hat{n}_{i_{}\uparrow}+
\hat{n}_{i_{}\downarrow}-1)}$) \cite{alba1,magalhaesphysicaa}.
This procedure allows to write the partition function   
for $2S$ and $4S$ situations in a compact form as 
\begin{equation}
Z\{y\}=e^{\frac{s-2}{2}\beta\mu}
\prod_{j_{}}\frac{1}{2\pi}
\displaystyle \int_{0}^{2\pi} dx_{j_{}}e^{-y_{j_{}}}~Z_{eff}\{y\}
\label{eqz}
\end{equation} 
and
\begin{equation}
Z_{eff}\{y\}=\int D(\varphi^{*}\varphi)e^{\left( A\{y_{j_{}}\}\right)}
\label{eqz1}
\end{equation} 
where $s$ ($s=2$ or $4$) is the number of states per site,
\begin{equation}
\begin{split}
A\{y_{j_{}}\}&=\int^{\beta}_{0}d\tau \left\lbrace  \sum_{j,\sigma}\varphi^{*}_{j\sigma}(\tau)
\left[ -\frac{\partial}{\partial\tau}   +  \frac{y_{j}}{\beta}\right]\varphi_{j\sigma}(\tau)
\right.  \\ &-   \left. H(\varphi^{*}_{j\sigma}(\tau),\varphi_{j\sigma}(\tau))  \right\rbrace 
\label{eqz2},
\end{split}
\end{equation} 
$\mu$ is chemical potential, $\beta=1/T$ and $y_{j}=\beta\mu$ or $y_{j}=ix_{j}$ for the
$4S$-FvH and $2S$-FvH models, respectively. 

The action $A\{y\}$ given in Eq. (\ref{eqz1}) can be build using the hamiltonian (\ref{ham}). 
Thus, we have: 
\begin{equation}
A\{y_{j_{}}\}=A_{\Gamma}+A_{SG}+A_{FE}
\label{acao}
\end{equation}
where
\begin{equation}
A_{\Gamma}=\int_{0}^{\beta}\sum_{j}\underline{\varphi}_{j}^{\dag}(\tau)
\left[\frac{y_j}{\beta}-\frac{\partial}{\partial \tau} +\Gamma\underline{\sigma}_{x}
\right]\underline{\varphi}_{j}(\tau),
\label{eqao}
\end{equation}
\begin{equation}
A_{SG}=\int_{0}^{\beta}\sum_{(i,j)} \frac{J_{ij}}{2}S_{i}(\tau)S_{j}(\tau),
\label{asg1}
\end{equation}
\begin{equation}
A_{FE}=\frac{J_{0}}{N}
\int_{0}^{\beta}d\tau \sum_{(i,j)}S_{i}(\tau)S_{j}(\tau)
\label{ferro}
\end{equation}
and
\begin{equation}
S_{i}(\tau)=\underline{\varphi}_{i}^{\dag}(\tau)
\underline{\sigma}^z \underline{\varphi}_{i}(\tau)
\label{eqs}.
\end{equation}
\noindent
The matrices in Eqs. (\ref{eqao})-(\ref{eqs}) are defined as:
\begin{equation}
\underline{\varphi}_{i}(\tau)=\left[\begin{tabular}{c}$\varphi_{i\uparrow}(\tau)$ \\ 
$ \varphi_{i\downarrow}(\tau)$ \end{tabular}
\right]; ~
\underline{\sigma}^x=\left(
\begin{tabular}{cc}
$0$ & $1$\\  $1$ & $0$
\end{tabular}
\right); 
\ \
\underline{\sigma}^z=\left(
\begin{tabular}{cc}
$1$ & $0$\\  $0$ & $-1$
\end{tabular}
\right).
\ \
\label{pauli}
\end{equation}
\noindent

The  coupling $J_{ij}$ given in Eq. (\ref{Jij}) allows to rewrite the random part of the 
action as:
\begin{equation}
\begin{split}
A_{SG}= 
\frac{J}{2N}\int_{0}^{\beta}d\tau\left\{
   \left[ {\displaystyle\sum_{j=1}^{N}}(\eta_{j}+\xi_{j})S_{j}(\tau)\right]^{2}
   -
  \left[  {\displaystyle\sum_{j=1}^{N}}\eta_{j}S_{j}(\tau)\right] ^2\right.
\\
-\left. \left[
{\displaystyle\sum_{j=1}^{N}}\xi_{j}S_{j}(\tau)\right] ^2
-2{\displaystyle\sum_{j=1}^{N}}\eta_{j}\xi_{j}(S_{j}(\tau))^2\right\}
\label{eq9}.
\end{split}
\end{equation}
Similarly, the ferromagnetic part of the 
action is given as
\begin{equation}
A_{FE}
=\frac{J_{0}}{2N}\int_{0}^{\beta} d\tau\left\{
\left[ {\displaystyle\sum_{j}}S_{j}(\tau)\right] ^2-
{\displaystyle\sum_{j}}\left[
S_{j}(\tau)\right]^2\right\}.
\label{eq101}
\end{equation}

The Hubbard-Stratonovich transformation is used
to linearize the first three terms and the first one from the right side of 
Eqs. (\ref{eq9}) 
and (\ref{eq101}), respectively. 
The last terms in the right side of 
Eqs. (\ref{eq9})-(\ref{eq101}) vanish in the thermodynamic limit.
Therefore, $Z\{y\}$ defined in Eqs. (\ref{eqz})-(\ref{eqs})
is obtained as:
\begin{equation}
\begin{split}
Z\{y\}=e^{\frac{s-2}{2}\beta\mu} 
{\displaystyle \int Dq_{1}(\tau) }{\displaystyle \int Dq_{2}(\tau)}
 {\displaystyle \int Dq_{3}(\tau)}{\displaystyle \int Dm(\tau)}
\\
\times
\exp \{ -N 
[ \frac{1}{2}\int_{0}^{\beta}d\tau (q_{1}(\tau)^{2}+ q_{2}(\tau)^{2}+ q_{3}(\tau)^{2} + m(\tau)^{2})
\\ 
  -\frac{1}{N} << \ln \left( \int_{0}^{2\pi}\prod_{j}\frac{dx_j}{2\pi} e^{-y_j}
\Lambda (q_{1},q_{2},q_{3}, m) \right)>>_{\eta\xi} ]\} 
\label{eq11}
\end{split}
\end{equation}
with
\begin{equation}
\begin{split}
\Lambda(q_{1}, q_{2},q_{3},m)= {\displaystyle \int
D(\varphi^{*}\varphi)}
\\ \times \exp \left(
{\displaystyle\sum_{j}}{\displaystyle\int_{0}^{\beta} d\tau\underline{\varphi}_{j
}^{\dag}(\tau) \underline{G}^{-1}(\tau|h_{j})\underline{\varphi}_{j
}(\tau)}\right) 
\label{eq19}
\end{split}
\end{equation}
and
\begin{equation}
 \underline{G}(\tau|h_{j})=
\left[
\frac{y_j}{\beta}-\frac{\partial}{\partial \tau}+h_{j}\underline{\sigma}^{z}
+\beta\Gamma\underline{\sigma}^{x}
\right] .
\label{eq20}
\end{equation}
In Eq. (\ref{eq11}), it has been used the self-averaging property
$\frac{1}{N}\sum_{j}f(\xi_{j};\eta_{j})$. The average $<<...>>_{\xi,\eta}$ can be achieved by using
the probability distribution $P(\xi,\eta)=P(\xi)P(\eta)$ where $P(\xi)$ (or $P(\eta)$) is defined in
Eq. (\ref{prob}).
In this work, the SA is assumed, in which $m=m(\tau)$ and $q_n=q_n(\tau)$ for $n=$1, 2 and 3.
Therefore,  the internal 
field in Eq.(\ref{eq20}) is
\begin{equation}
h_{j}
= \sqrt{J} \left(i\eta_{j} q_{1}+ i \xi_{j} q_{2}+(\eta_{j}+\xi_{j})q_{3}\right)
+ 
\sqrt{J_0}\ m.
\label{eq21}
\end{equation}

The integrals over \{$q_n$\} 
and $m$ in Eq. (\ref{eq11}) can be solved by the saddle point method 
which gives 
\begin{equation}
 q_1=i\ \sqrt{ J} \frac{1}{N}\sum_{j} \langle \xi_{j}{S}_{j}^{}\rangle=i\ \sqrt{
J}\ Q_1
\label{eqq_1}
\end{equation}
\begin{equation}
 q_2=i\ \sqrt{ J} \frac{1}{N}\sum_{j} \langle \eta_{j} {S}_{j}^{}\rangle =i\ \sqrt{ J}\
Q_2
\label{eqq_2}
\end{equation}
$q_3= Q_1+ Q_2$ and
\begin{equation}
 m= \sqrt{ J_0 } \frac{1}{N}\sum_{j} \langle {S}_{j}^{}\rangle =\sqrt{ J_0}\ M.
\label{eqq_m}
\end{equation}
\begin{figure*}[htb]
\includegraphics[angle=270,width=14cm]{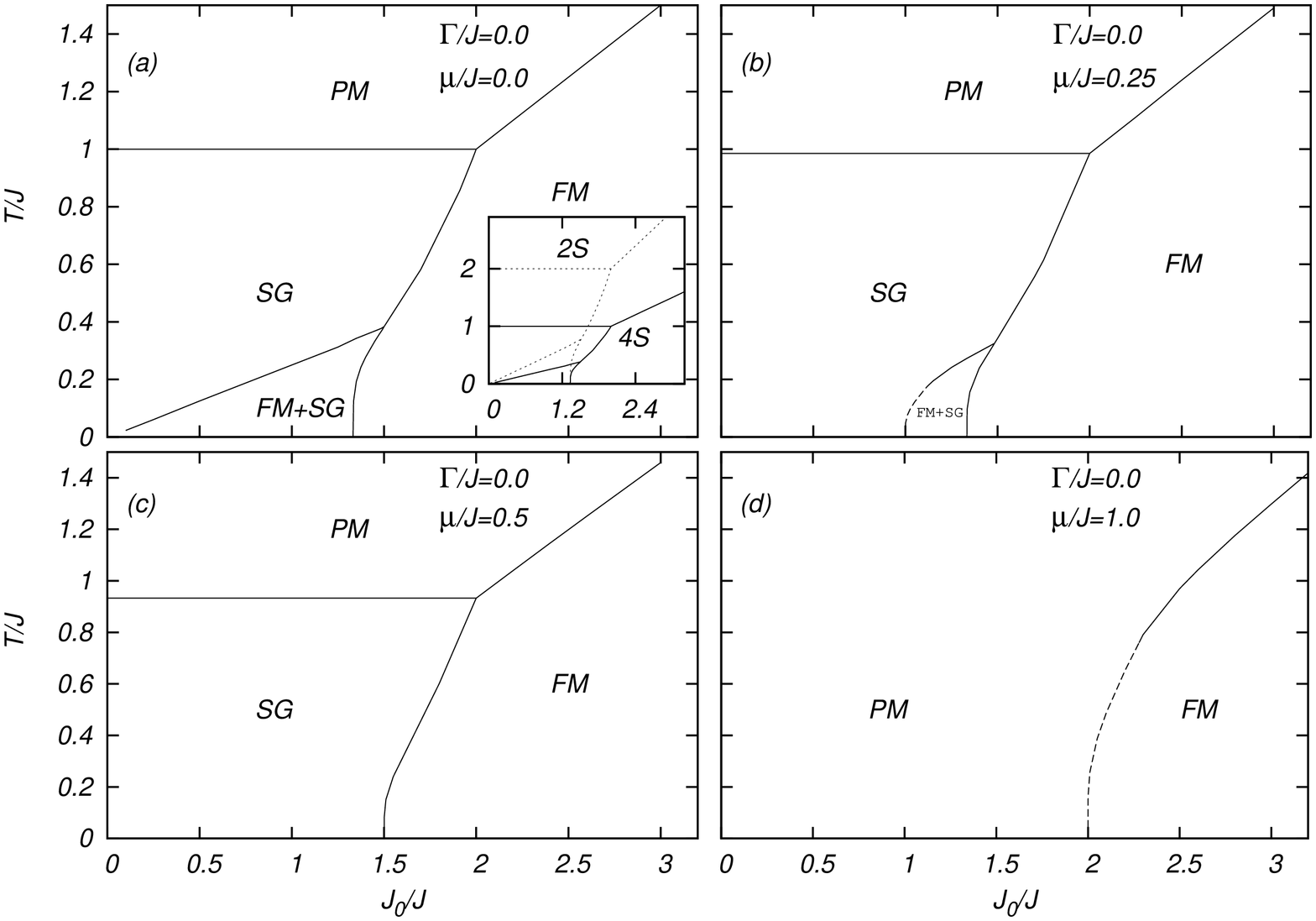}
\centering
\caption{Phase diagrams $T/J$ versus $J_0/J$ for several values of $\mu/J$ and $\Gamma/J=0$.
The solid lines correspond to the second order transition while the dashed lines indicate
first order transition. The inset shows a comparison between the phase diagrams for the
$2S$-fvH model (dotted lines) and the $4S$-fvH model with $\mu=0$.
}\label{fig1}
\end{figure*}
The Fourier transform can be used in Eq. (\ref{eq19}) and the functional integrals over the
Grassmann fields ($\varphi,~\varphi^{*}$) 
performed as well as the subsequent sum over Matsubara's frequencies 
following close Ref. \cite{albaphysicaa} to give:
\begin{equation}
\Lambda(Q_{1}, Q_{2},M)=2 e^{y}[\cosh(y)+\cosh(\sqrt{\beta \Delta})] 
\label{Lambda}
\end{equation}
with
$\Delta=[J(\eta +\xi)(Q_1+Q_2)-J\eta Q_1-J\xi Q_2+
J_0 M]^{2} + \Gamma^{2}$.

For the $2S$-FvH model the integral over $x$ in Eq. (\ref{eq11}) 
is calculated which leads to
the same thermodynamics 
found in Ref. \cite{jricardo1}  for the quantum vH model.
While, in the case of 
$4S$-FvH model, $Z\{\mu\}$ can also be obtained from Eq.(\ref{eq11}) which gives for the grand
canonical potential  $\Omega$, the 
following 
expression: 
\begin{equation}
\begin{split}
\beta\Omega = \beta J Q^{2} + \frac{\beta J_0}{2} M^{2} - \beta\mu 
\\
- \langle\langle \ln
2[\cosh(\beta\mu) + \cosh(\beta\sqrt{\Delta})]\rangle\rangle_{\eta\xi}
\label{potential}
\end{split}
\end{equation}
where it is assumed $Q_1=Q_2=Q$.

\begin{figure*}[htb]
\includegraphics[angle=270,width=14cm]{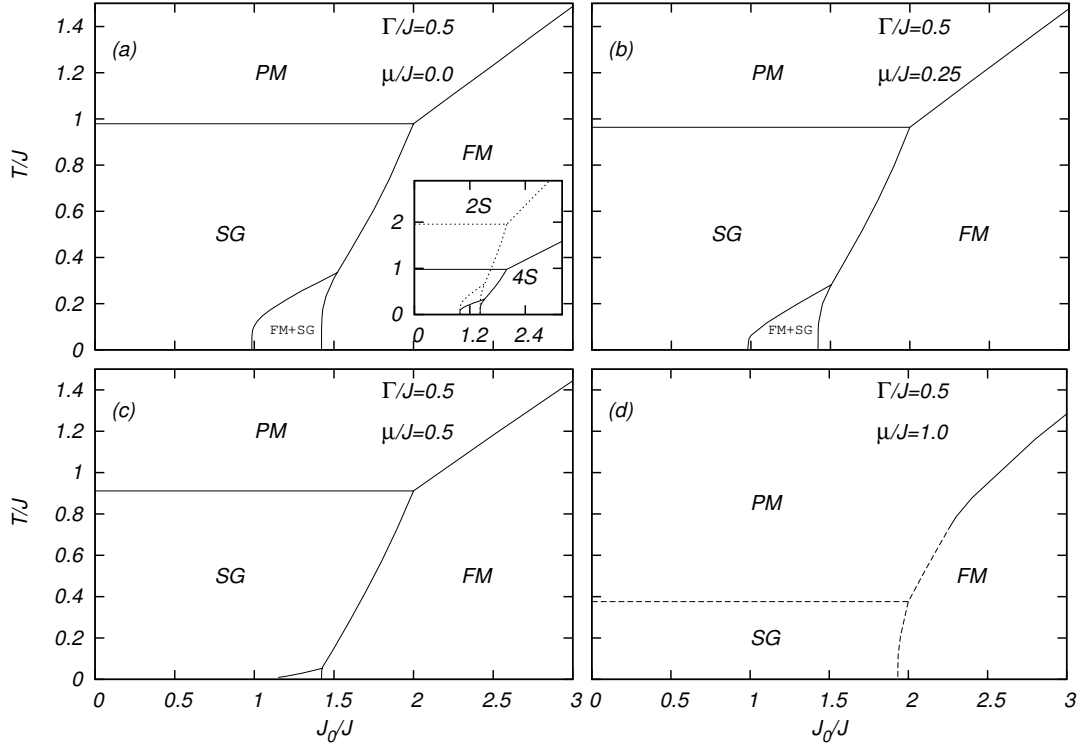}
\centering
\caption{Phase diagrams $T/J$ versus $J_0/J$ for several values of $\mu/J$ and $\Gamma/J=0$.
The solid lines correspond to the second order transition while the dashed lines indicate
first order transition. The inset shows a comparison between the phase diagrams for the
$2S$-fvH model (dotted lines) and the $4S$-fvH model with $\mu=0$.
}\label{fig2}
\end{figure*}

\begin{figure*}[htb]
\centering
\includegraphics[angle=270,width=14cm]{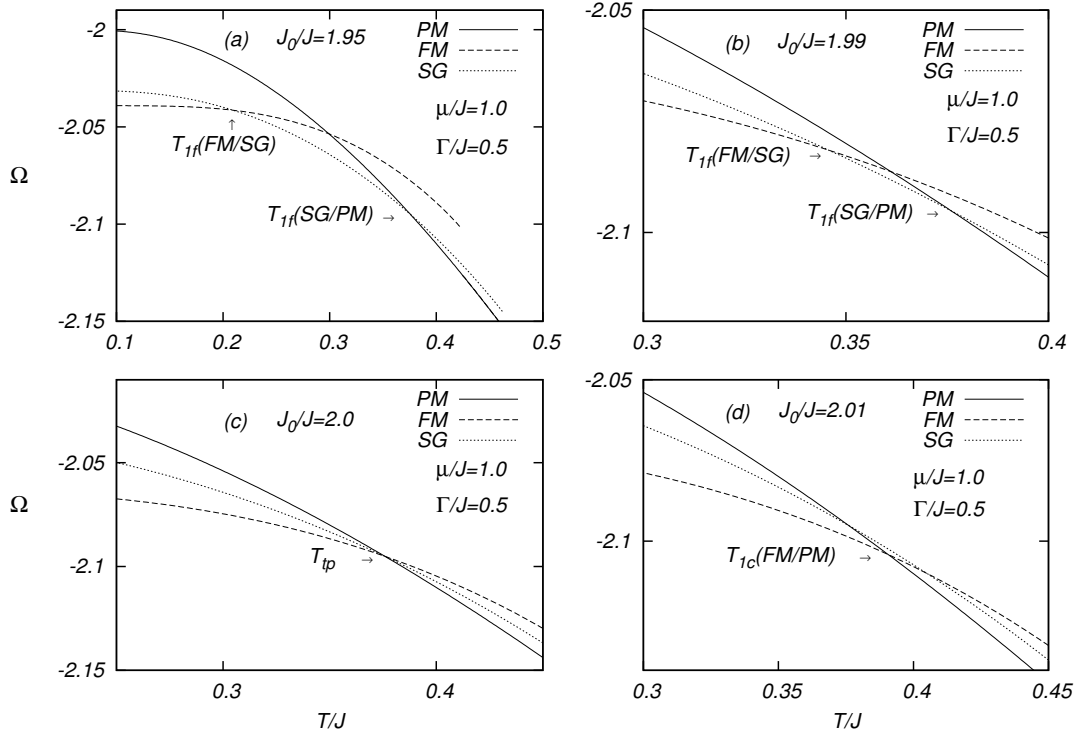}
\caption{Grand canonical potential $\Omega$ versus $T/J$ for $\mu/J=1.0$, $\Gamma/J=0.5$ and
several values of $J_0$.
}\label{figpot}
\end{figure*}
\begin{figure*}[htb]\includegraphics[angle=270,width=14cm]{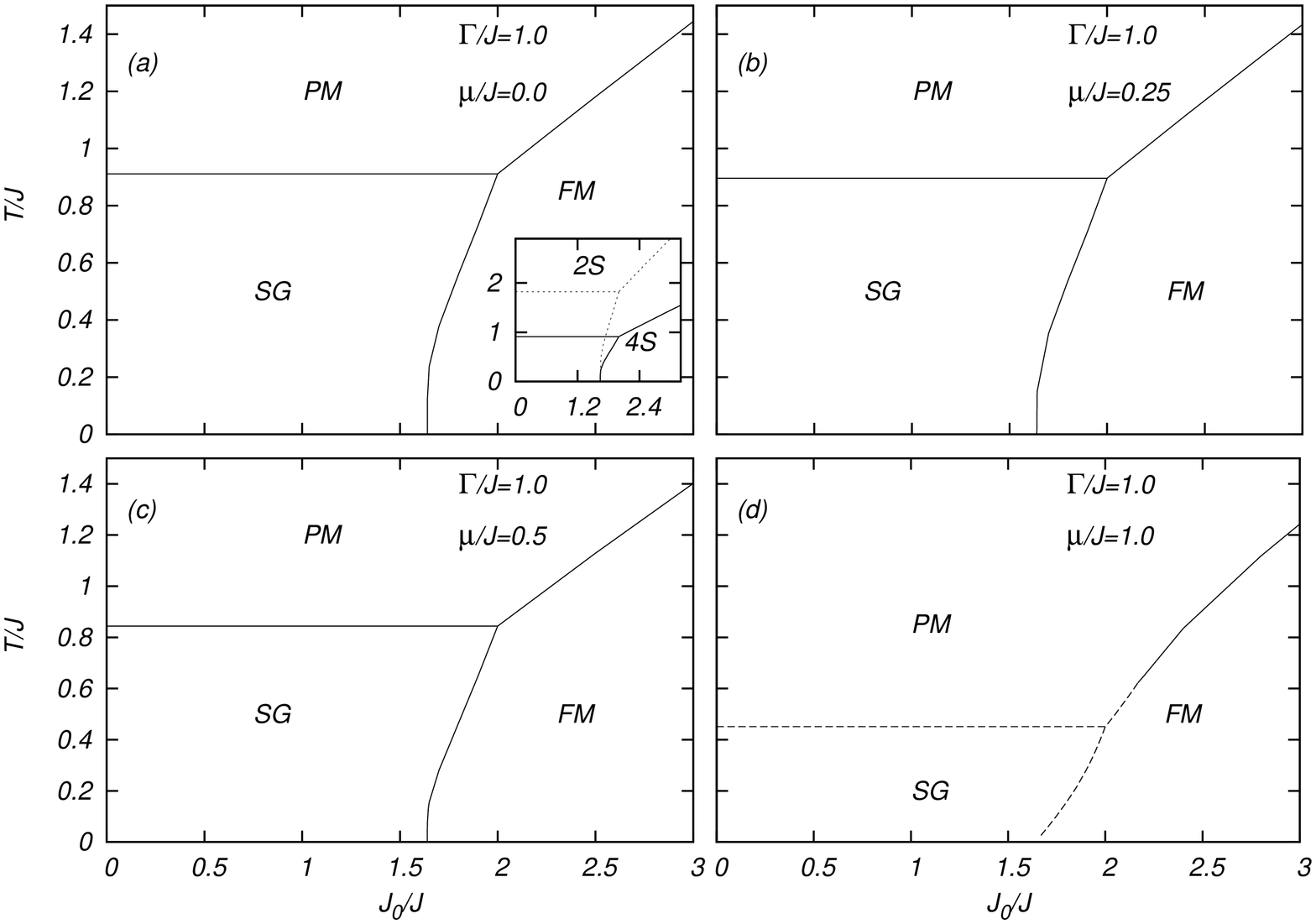}
\centering
\caption{Phase diagrams $T/J$ versus $J_0/J$ for several values of $\mu/J$ and
$\Gamma/J=1.0$.
 Dashed lines indicate first order transition. The inset shows phase diagrams for the $2S$-fvH
model (dotted lines) and the $4S$-fvH model with
$\mu=0$ and $\Gamma=1$.
}\label{fig23}
\end{figure*}

\section{Results}

In order to investigate the interplay between charge and spin fluctuations in the $4S$-FvH model in
the presence of a transverse magnetic field $\Gamma$, a set of phase diagrams can be built from the
numerical solution of the saddle point SG order parameter $Q$ and the magnetization $M$ which is
obtained from Eq. (\ref{potential}). It should be remarked that temperature, $J_0$ (the strength of
the ferromagnetic component of $J_{ij}$), $\mu$ and $\Gamma$ are given in units of $J$ (the strength
of the random component of $J_{ij}$). For numerical results, $J=2$ is used. In Figs.
\ref{fig1}-\ref{fig23}, phase diagrams temperature versus $J_0$ for several values of $\mu$ and 
$\Gamma$ are displayed.

Fig. \ref{fig1}(a) shows the situation corresponding to the half-filling occupation ($\mu=0$) and
$\Gamma=0$. For that case, the phase diagram of the classical vH model is basically recovered. For
small $J_0$, when the temperature is decreased, there is first a transition from the PM phase to the
SG one which has $M=0$ and $Q\neq 0$. Then, there is a second transition from the SG phase to the
mixed (FM+SG) one in which $M\neq 0$ and $Q\neq 0$. For larger $J_0$, SG and FM+SG phases are
replaced by the ferromagnetic (FM) one with $M\neq 0$ and $Q=0$. In the inset of Fig. 
\ref{fig1}(a), the  $4S$-FvH model phase diagram with $\mu=0$ is compared to the $2S$-FvH one. These
two models present qualitatively the same results. However, the phase boundaries of the $4S$-FvH
model at the half-filling appear at lower temperatures than the $2S$-FvH model. That is a
consequence of the non-magnetic states allowed in the $4S$-FvH model. Figs.
\ref{fig1}(b)-\ref{fig1}(d) show  how the increase of $\mu$ can  affect phase boundaries even when
$\Gamma=0$. When the average occupation $n$ moves from the half-filling ($\mu=0$) to doubly occupied
sites ($\mu=1$), the transition temperature $T_f$ between paramagnetism and SG phase as well as that
one (here called $T_g$) between SG and FM+SG phases are depressed. Ultimately, this depressing of
$T_f$ and $T_g$ as $\mu$ increases leads to the complete suppression of SG and FM+SG. Similarly to
the FISG model (see discussion in Ref. \cite{oppermann2}), the increase of $\mu$ also changes the
nature of the phase transitions which appear in Fig. \ref{fig1}. That can be seen, for instance, for
$T_{g}$ in Fig. \ref{fig1}(b) when $\mu=0.25$. Furthermore, for $\mu=1$ (see Fig. \ref{fig1}(d)),
the nature of the Curie temperature $T_c$ is also changed. It appears a tricritical point which can
be located (see Appendix) at $T_{c}^{tric}=0.759$ with $J_{0}^{tric}=2.278$. One last consequence
due to the increase of $\mu$ is that larger values of $J_0$ are necessary to obtain a FM solution.

In Fig \ref{fig2}, $\Gamma=0.5$ while $\mu$ assumes the same values already used in Fig. \ref{fig1}.
Therefore, the results can be compared directly to Fig. \ref{fig1} which allows to understand the
combined effects of $\Gamma$ and $\mu$. The phase diagram for $\mu=0$ is shown in Fig.
\ref{fig2}(a). In this case, the effect of $\Gamma$ is to depress both $T_{f}$ and $T_{g}$. However,
the subsequent increase of $\mu$ showed in Figs. \ref{fig2}(b)-\ref{fig2}(d) leads to a distinct
situation as compared with Fig \ref{fig1}. For instance, the FM+SG phase is completely suppressed.
In contrast, the SG phase is still preserved. Other important difference in comparison to the case
displayed in Fig. \ref{fig1} is concerned with the nature of phase boundaries. For instance, the
line transition $T_g$ remains always a second order one. That is not the case for $T_f$ which
becomes a first order one as displayed in Fig. \ref{fig2}(d) where $\mu=1$. In fact, for this value
of chemical potential there is a completely new scenario. The existence of the tricritical point in
the Curie temperature is preserved being located at $T_{c}^{tric}=0.729$. Nevertheless, there is now
an additional triple point $T_{triple}\approx 0.376$ with the coexistence of SG, FM and PM phases.
Fig. \ref{figpot} exhibits the grand canonical potential $\Omega$ as a function of temperature for
the SG, PM and FE phases near the triple point. The first order transitions are located by comparing
$\Omega$ of these phases, where the stable solution is always that one with minimum $\Omega$.
Particularly, the triple point is obtained when $\Omega$ of the SG, PM and FE phases assume the same
values (see Fig. \ref{figpot}(c)) 

In Fig. \ref{fig23} phase diagrams with $\Gamma=1.0$ are 
displayed.
In that case, the FM+SG is
completely suppressed for $\mu=0$ (see Fig. \ref{fig23}(a)).  The remaining phase diagrams 
displayed
in
Figs. \ref{fig23}(b) -\ref{fig23}(d) present only SG, FM and PM solutions for the order parameters.
When $\mu=0.25$ and $0.5$, the second order line transition $T_{f}$  is even more diminished when
compared to the respective previous situations given in Figs. \ref{fig1}-\ref{fig2}. For $\mu=1$,
the complex scenario found in Fig. \ref{fig2}(d) is repeated in Fig. \ref{fig23}(d) with the
presence of tricritical and triple points. However, interestingly, $T_f$, which is now a first order
line transition,  has a small increase as compared with Fig \ref{fig2}(d).

The effects of the $\Gamma$ increasing on the $2S$-FvH model can be observed in the inset of Figs.
\ref{fig2}(a) and \ref{fig23}(a), which exhibit phase diagrams of the $2S$-FvH model compared to
the $4S$ one with $\mu=0$ for $\Gamma=0.5$ and 1.0, respectively. As in the $4S$-FvH model,
the magnetic phase is also suppressed by $\Gamma$ in the $2S$-FvH model. In addition, the phase
boundaries converge to the same values when temperature decreases.

\section{Conclusion}

The present work has studied the fermionic version of the classical van Hemmen model \cite{vh} for
SG in the presence of a transverse magnetic field $\Gamma$. The goal is to verify how the interplay
between charge and spin fluctuations can affect the magnetic ordering in this particular model. It
should be mentioned that the knowledge of such interplay can also be quite useful in further
applications for studying the disordered magnetism when other physical processes are present (see,
for instance, Ref. \cite{magalhaesprb2}). 
 
The mentioned fermionic version has been formulated in two versions. In the first one, the so called
$2S$-FvH, the two nonmagnetic eigenstates of the operator $\hat{S}_{i}^{z}$ (defined in Eq.
(\ref{eqs}) have been eliminated. Therefore, the corresponding thermodynamics obtained is exactly
the same one obtained by a previous quantum formulation of the van Hemmen model given in Ref.
\cite{jricardo1}. On the contrary, when the four original eigenstates are preserved, the solution
for both saddle point order parameters, the SG ($Q$) and the magnetization ($M$) one, indicates that
not only the phase diagram but also the nature of the phase transitions are deeply affected when the
chemical potential $\mu$ and/or $\Gamma$ are varied. For instance, the mixed phase FM+SG can be
suppressed as well as there is onset of tricritical and triple points by increasing $\mu$ and/or
$\Gamma$. In fact, the role of $\Gamma$ as responsible for the suppression of the FM+SG phase has
already been found in Ref. \cite{jricardo1}. Remarkably, in the present problem, the same result can
be obtained by varying only $\mu$, or in other words, only by changing the average occupation of
sites $n$. Therefore, one can say that the thermodynamics derived from van Hemmen model is either
sensitive to a spin flipping mechanism provided by $\Gamma$ as well a dilution one provided by
$\mu$. Surely, when both mechanisms are combined, the complexity of the phase diagram is even more
intense as can be illustrated by a first order phase transition between the PM and SG (or FE) phases
with the presence of tricritical and triple points. 

To conclude, we have shown in this work that mechanisms such as dilution and quantum spin flipping
given by $\mu$ and $\Gamma$, respectively, are sources of quite non-trivial effects in the fermionic
van Hemmen model. More precisely, those effects indicate that the existence of magnetic phases and
the nature of its boundaries in that model are extremely sensitive to the redistribution of charge
provided by $\mu$ and $\Gamma$. It should be emphasized that thermodynamics in the present work
could be derived without the use of sophisticated mathematical techniques such as the replica
method. It is well known that in the case of SG first order phase transitions, the use of this
method implies in considerable difficulties  as, for example, location of first order phases
boundary \cite{magalhaesprb3, magalhaesprb4, ghatak, salinas, mottishaw1, mottishaw2}.

\section*{Acknowledgments}
This work was partially supported by the Brazilian agencies CNPq  and CAPES.

\appendix
\section{}\label{}
The Landau expansion of the grand canonical potential (Eq. (\ref{potential})) in powers
of $M$ and $Q$ is given by
\begin{equation}
 \beta \Omega = \Omega_0+A_2 Q^2 +\frac{\beta^2 J^4 A}{\Gamma^2} Q^4 + B_2 M^2 +\frac{\beta^2
J_0^4 A}{8\Gamma^2} M^4
\end{equation}
where
\begin{equation}
 A_2=\beta J(1-\frac{J\sinh \beta\Gamma}{\Gamma K_0}),\  B_2=\frac{\beta J_0}{2}(1-\frac{J_0\sinh
\beta\Gamma}{\Gamma K_0}),
\end{equation}
\begin{equation}
 A=-\frac{1+\cosh\beta\Gamma\cosh\beta\mu}{K_0^2}+\frac{\sinh \beta\Gamma}{\beta\Gamma K_0}
\end{equation}
and
\begin{equation}
K_0=
\cosh\beta\mu+\cosh{\beta\Gamma} .
\end{equation}
The second order phase transition from the paramagnetic phase to the SG one (to the FE phase) 
occurs when $A_2=0$ ($B_2=0$) and $A>0$. The tricritical point is located when $B_2=0$ with $A=0$. 





\end{document}